\documentclass[twocolumn,secnumarabic,amssymb, nobibnotes, aps, prd]{revtex4-1}

\begin{document}
\title{\bf Braneworld teleparallel gravity }
\author{K. Nozari}%
\email{knozari@umz.ac.ir}
\author{A. Behboodi}%
\email{a.behboodi@stu.umz.ac.ir}
\author{S. Akhshabi}%
\email{s.akhshabi@umz.ac.ir}

\affiliation{Department of Physics, Faculty of Basic Sciences,
University of Mazandaran, P. O. Box 47416-95447, Babolsar, IRAN}
\vspace{1cm}

\begin{abstract}
We study the gravity in the context of a braneworld teleparallel
scenario. The geometrical setup is assumed to be Randall-Sundrum II
model where a single positive tension brane is embedded in an
infinite AdS bulk. We derive the equivalent of Gauss-Codacci
equations in teleparallel gravity and junction conditions in this
setup. Using these results we derive the induced teleparallel field
equations on the brane. We show that compared to general relativity,
the induced field equations in teleparallel gravity contain two
extra terms arising from the extra degrees of freedom in the
teleparallel Lagrangian. The term carrying the effects of the bulk
to the brane is also calculated and its implications are discussed.
\begin{description}
\item[PACS numbers]
04.50.Kd
\item[Key Words]
Teleparallel Gravity, Braneworld Gravity
\end{description}
\end{abstract}
\vspace{2cm}
\maketitle
\section{Introduction}
The theory of general relativity (GR) first introduced by Einstein
in 1916, is one of the great achievements of theoretical physics.
The remarkable agreement of GR with experimental data, at least in
the solar system scales, turns it into one of the most successful
theories ever developed. However there are some dark spots that it
is not able to answer. For example, the theory of general relativity
effectively predicts its own demise with the prediction of
singularities. Also the problem of combining gravity with quantum
theory and the issue of unification of fundamental forces, make
modifying GR a more plausible task. It is obvious that in extremely
high energy levels (Planck scale) the theory of GR breaks down,
hence the idea of modifying gravity came to existence. One of these
ideas is higher dimensional world theories which originated by the
work of Kaluza and Klein \cite{KK}. They considered the extra
dimension in their attempt to unify electromagnetism with gravity.
In 1998, Arkani-Hamed, Dimopoulos and Dvali (ADD) showed that the
extra dimension can be in order of $0.1$mm \cite{ADD}. The extra
dimensions of the ADD scenario are flat and the model postulates
that the Standard Model fields are confined to a 4-dimensional
brane, with only gravity propagating in the bulk. In the two
Randall-Sundrum (RS) models proposed shortly afterwards, the extra
dimension is not flat and the bulk geometry is curved. This leads
the brane(s) to have tension so the brane(s) and the bulk can
dynamically interact with each other. Similar to the ADD scenario,
in the first Randall-Sundrum model (RS I) the extra dimension is
still compact \cite{RS1}. The RS I model consists of two branes of
tensions $\lambda_{1}$ and $\lambda_{2}$ bounding a slice of anti-de
Sitter (AdS) space. The main aim of the first RS model was to solve
the Hierarchy problem. The inter-brane separation in this model is
characterized by an extra degree of freedom called Radion. To get
the desired result of the theory the Radion should be stabilized
which brings up some complexities \cite{GW}. Moreover, the model
leads to unconventional and unacceptable cosmology in the weak-field
limit on the brane. In the second RS model (RS II), the extra
dimension is infinite in size and the bulk curvature causes the
gravity to be localized on the brane \cite{RS2}. The RS II model
consists of a single, positive tension brane in an infinite bulk.
The model can be thought of as a RS I model when the negative
tension brane is moved to infinity. This model does not address the
hierarchy problem, but has some remarkable gravitational and
cosmological implications. The elegance and geometrical simplicity
of this model had a huge impact on studies of extra dimensions and
led to vast literature researching various aspects of it in gravity
\cite{RSG} and cosmology \cite{RSC}. This is the model we use as our
geometric framework in this paper. We follow the procedure
introduced by Shirumizu, Maeda and Sasaki in 2000 \cite{SMS}. They
obtained the 4-Dimensional field equations on the brane by a
geometrical approach projecting the 5-D quantities to the brane.

On the other hand, in 1928,  along with general relativity Einstein
presented another form of theory of gravity called teleparallel
gravity \cite{Ein30}. In this theory, a set of four tetrad (or
vierbein) fields form the (psoudo)-orthogonal bases for the tangent
space at each point of spacetime and the torsion instead of
curvature describes gravitational interactions. Tetrads are the
dynamical variables and play a similar role to the metric tensor
field in general relativity. Teleparallel gravity also uses the
curvature-free Weitzenb\"{o}ck connection instead of Levi-Civita
connection of general relativity to define covariant derivatives
\cite{Weitz23}.

Since the first introduction, some innovative works developed the
theory and it has been shown that teleparallel Lagrangian density
only differs with Ricci scalar by a total divergence
\cite{Haya79,Ald10}. This shows that general relativity and
teleparallel gravity are dynamically equivalent theories where the
difference arises only in boundary terms. However, there are some
fundamental conceptual differences between teleparallel theory and
general relativity. According to general relativity, gravity curves
the spacetime and shapes the geometry. In teleparallel theory
however torsion does not shape the geometry but instead acts as a
force. This means that there are no geodesics equations in
teleparallel gravity but there are force equations much like the
Lorentz force in electrodynamics. In teleparallel gravity one would
consider the nontrivial characteristics of spacetime in the (pseudo)
orthogonal bases of the tangent space (tetrad) and the spacetime is
considered flat. The relation between manifold and Minkowski metrics
is
\begin{equation}
g_{\mu\nu}=\eta_{ij}e_{\mu}^{i}e_{\nu}^{j}
\end{equation}
The Greek indices referred to coordinate basis of the manifold and
Latin indices referred to basis of the tangent space. Both indices
run from $0$ to the dimension of the spacetime. Only the spin
connection $A^{a}_{bc}$ acts on tangent space indices and the
Weitzenb\"{o}ck connection $\Gamma^{\rho}_{\mu\nu}$ only acts on
spacetime indices. The spin connection in teleparallel gravity is
only related to Weitzenb\"{o}ck connection like general relativity
which is only related to Livi-Civita connection. To satisfy the
absolute parallelism condition in teleparallel gravity, the spin
connection is supposed to be vanished. This leads to lack of spinor
fields which guarantees the equivalence of the two theories. One can
define the curvature and torsion with respect to spin connection and
vierbeins \cite{Ald10} which in teleparallel gravity leads to zero
curvature and non-zero torsion as
\begin{equation}
T^{\rho}_{\,\,\,\,\mu\nu}\equiv
e_{i}^{\rho}(\partial_{\mu}e_{\nu}^{i}-\partial_{\nu}e_{\mu}^{i})
\end{equation}
In attempts to obtain the accelerated expansion of the universe the
extended version of the theory has been presented. In $f(T)$
gravity, in analogy with $f(R)$ the higher order terms of $T$ is
considered in the Lagrangian of teleparallel action. In the case of
$f(T)=T$ the Lagrangian can be assumed to be locally Lorentz
invariant. Nevertheless, for $f(T)\neq T$ the theory is not locally
Lorentz invariant anymore. This property would bring more degrees of
freedom to the theory in comparison with $f(R)$ gravity\cite{LI}.
For more information about the theory and its extended versions see
\cite{FT} and references therein. Notwithstanding the equivalence of
the two theories, in the situations where boundary and induced terms
are present, teleparallel gravity would not be equal to general
relativity anymore. The reason for this is that for a given
``induced'' metric, there exist an infinite number of ``induced''
vierbeins that give the same metric and each of these vierbeins
results in different field equations. Such boundary terms exist in
brane world theories which brings up the questions about the
interpretation of gravity in higher dimensional teleparallel
scenarios. We wish to answer some of these questions in this paper,
using the RS II model as our braneworld geometrical setup.

The structure of the paper is as follows: In section II we review
basic definitions of teleparallel gravity and set out the notation
for the rest of the paper. In section III,  we derive the equivalent
of Gauss-Codacci equations in teleparallel gravity using our
definition of projection vierbein. In section IV, the junction
conditions in our model is obtained. In section V, using the results
of previous sections and following the procedure of reference
\cite{SMS}, we derive the induced teleparallel field equations on
the brane and finally in section VI conclusion and some discussions
are presented.

\section{Notation and Definitions}
In brane teleparallel gravity we define five tetrad fields
$\textbf{e}_{i}(x^{\mu})$ which form the basis of tangent space at
each point of the manifold with spacetime coordinates $x^{\mu}$:
$\textbf{e}_{i}\,.\,\textbf{e}_{j}=\eta_{ij}$. Latin indices labels
the tangent space coordinates while Greek indices label spacetime
coordinates. Both set of indices take the values $0,1,2,3,5$ to
agree with conventional notation. $\textbf{e}_{i}(x^{\mu})$ is a
vector in the tangent space and can be expressed in terms of its
components in the coordinates space as $e_{i}^{\mu}$ which is called
the vierbein. Vierbeins are also vectors in spacetime. The relation
between the Minkowski and the spacetime metrics is as mentiond in
equation (1). The inverse vierbein is defined by the relation
$e_{i}^{\mu}e_{\mu}^{j}=\delta_{i}^{j}$. The vierbein is also a
matrix that transforms between the tetrad frame and the coordinate
frame.\\
The connection in teleparallel theory, the Weitzenb\"{o}ck
connection, is defined as
\begin{equation}
\Gamma^{\rho}_{\,\,\,\,\mu\nu}=e^{\rho}_{i}\partial_{\nu}e^{i}_{\mu}
\end{equation}
By this definition, the torsion tensor and its permutations are
\cite{Haya79}
\begin{equation}
T^{\rho}_{\,\,\,\,\mu\nu}\equiv
e_{i}^{\rho}(\partial_{\mu}e_{\nu}^{i}-\partial_{\nu}e_{\mu}^{i})
\end{equation}
\begin{equation}
K^{\mu\nu}_{\quad\rho}=-\frac{1}{2}(T^{\mu\nu}_{\quad\rho}
-T^{\nu\mu}_{\quad\rho}-T_{\rho}^{\,\,\,\,\mu\nu})
\end{equation}
\begin{equation}
S^{\,\,\,\,\mu\nu}_{\rho}=\frac{1}{2}(K^{\mu\nu}_{\quad\rho}
+\delta^{\mu}_{\rho}T^{\alpha\nu}_{\quad\alpha}-\delta^{\nu}_{\rho}
T^{\alpha\mu}_{\quad\alpha})\,,
\end{equation}
where $K^{\mu\nu}_{\quad\rho}$ is the contortion tensor which is the
difference between Levi-Civita and the Weitzenb\"{o}ck connections
and $S^{\,\,\,\,\mu\nu}_{\rho}$ is called the superpotential. It
should be noted that the superpotential tensor carries all the extra
degrees of freedom in teleparallel gravity. In fact, due to
(anti-)symmetrical properties of the Weitzenb\"{o}ck connection,
compared to general relativity, there are more possible contractions
that can be used to define the Lagrangian of the theory. In
correspondence with Ricci scalar we define a torsion scalar as
\begin{equation}
T=S^{\,\,\,\,\mu\nu}_{\rho}T^{\rho}_{\,\,\,\,\mu\nu}
\end{equation}
so the gravitational action is
\begin{equation}
I=\frac{1}{16\pi G}\int d^{5}x\,|e|\, T
\end{equation}
where $|e|$ is the determinant of the vierbein $e^{a}_{\mu}$ which
is equal to $\sqrt{-g}$. Variation of the above action with respect
to the vierbeins will give the (5-dimensional) teleparallel field
equations
\begin{equation}
e^{-1}\partial_{\mu}(ee_{i}^{\rho}S^{\,\,\,\,\mu\nu}_{\rho})
-e_{i}^{\lambda}T_{\,\,\,\,\mu\lambda}^{\rho}S^{\,\,\,\,\nu\mu}_{\rho}
+\frac{1}{4}e_{i}^{\nu}T=4\pi G e_{i}^{\rho}\Xi^{\,\,\nu}_{\rho}
\end{equation}
where $\Xi^{\,\,\nu}_{\rho}$ is the 5-dimensional matter
energy-momentum tensor. Since in the Randall-sundrum II setup, the
bulk is assumed to be empty except for a cosmological constant and
the matter is assumed to be confined to the brane, we can decompose
the 5-dimensional stress-energy tensor as
\begin{equation}
\Xi_{\mu\nu}=-\Lambda_{5}g_{\mu\nu}+\delta(y)\frac{8\pi}{M^{3}_{5}}[-\lambda
g_{\mu\nu}+\tau_{\mu\nu}]
\end{equation}
where $\Lambda_{5}$ is the 5-dimensional cosmological constant,
$\lambda$ is the brane tension and $\tau_{\mu\nu}$ is the matter
stress-energy tensor of the brane which is assumed to be located at
$y=0$.

\section{Induced Quantities on the Brane: Gauss-Codacci equations}

Randall-Sundrum II geometry consists of a 4-dimensional hypersurface
or brane embedded in a 5-dimensional AdS bulk. In order to derive
the induced teleparallel field equations on the brane, it is
necessary to know how the 5-dimensional quantities can be projected
into the 4-dimensional hypersurface (brane). The projection tensor
on the brane, $q_{\mu\nu}$ is defined as
\begin{equation}
q_{\mu\nu}=g_{\mu\nu}-n_{\mu}n_{\nu}
\end{equation}
where $n_{\mu}$ is the unit normal vector on the brane. With this definition we
can also define a projection vierbein
\begin{equation}
h^{\mu}_{i}=e^{\mu}_{i}-n^{\mu}n_{i}\,.
\end{equation}
One should note that the projection vierbein defined here is not
unique. One can define other vierbeins to give the same induced
metric (11). $h^{\mu}_{i}$ acts as a projection operator on the
hypersurface as one can easily check. Any vector $v^{i}$ can be
decomposed into parts tangent and perpendicular to the hypersurface
$v^{i}=v_{\|}^{i}+v_{\bot}n^{i}$. Acting the projection vierbein on
this, we get
\begin{equation}
h_{i}^{\mu}v^{i}=(e^{\mu}_{i}-n^{\mu}n_{i})(v_{\|}^{i}+v_{\bot}n^{i})=v_{\|}^{\mu}
\end{equation}
so the projection vierbein projects the vector to the hypersurface
and also changes indices from coordinate to tangent and vice-versa.
Using the above equation we can define a covariant derivative
operator on the brane, $D_{\mu}$ by projecting all the indices in
the 5-dimensional covariant derivative using $h^{\mu}_{i}$
\begin{equation}
D_{\mu}T_{\,\,\,\,\,\,\beta_{1}...\beta_{ m}}^{\alpha_{1}...\alpha_{
n}}=h_{i_{1}}^{\alpha_{1}}...h_{i_{n}}^{\alpha_{n}}
h_{\beta_{1}}^{j_{1}}...h_{\beta_{n}}^{j_{n}}h^{k}_{\mu}\nabla_{k}T_{\,\,\,\,\,\,j_{1}...j_{
m}}^{i_{1}...i_{ n}}
\end{equation}
In order to derive the equivalent of the Gauss-Codacci
equations in teleparallel gravity, we start
with the following general equation. For any vector $X^{a}$ we have
\begin{equation}
\nabla_{d}\nabla_{c}X^{a}-\nabla_{c}\nabla_{d}X^{a}=R_{\,\,bcd}^{a}X^{b}+T_{\,\,cd}^{e}\nabla_{e}X^{a}
\end{equation}
In teleparallel gravity, where the connection is Weitzenb\"{o}ck,
the Riemann tensor identically vanishes, so we have
\begin{equation}
\nabla_{d}\nabla_{c}X^{a}-\nabla_{c}\nabla_{d}X^{a}=T_{\,\,cd}^{e}\nabla_{e}X^{a}
\end{equation}
where $\nabla$ is Weitzenb\"{o}ck covariant derivative here. Using
the above equation we can write
\begin{equation}
D_{\mu}D_{\nu}X_{\rho}-D_{\nu}D_{\mu}X_{\rho}=^{(N-1)}T^{\alpha}_{\,\,\,\mu\nu}D_{\alpha}X_{\rho}
\end{equation}
where $N$ is the dimension of the spacetime (5 in Randall-Sundrum
setup) and $D_{\mu}$ is the covariant derivative on the brane.
Substituting (14) in equation (17) and also considering the fact
that Weitzenb\"{o}ck covariant derivative of the tetrad vanishes, we
arrive at the teleparallel equivalent of the Gauss equation after
some straightforward algebra
\begin{equation}
^{(N-1)}T^{\rho}_{\,\,\,\mu\nu}=h_{k}^{\rho}h_{\nu}^{j}h_{\mu}^{i}\,
^{(N)}T^{k}_{\,\,\,ij}
\end{equation}
The interesting point is that unlike general relativity where the
extrinsic curvature explicitly enters the equation,  the ``extrinsic
torsion'' is not present in the R.H.S of the Gauss equation in
teleparallel gravity. The reason lies in the form of equation (16).
In general relativity the arbitrary vector appears in the R.H.S but
in teleparallel gravity covariant derivative of the vector appears.
Substituting 4-dimensional covariant derivative with 5-dimensional
ones results in term involving the extrinsic torsion to drop out of
the final result.

Starting with (18), we can now project all the terms in the L.H.S
side of the teleparallel field equation on the brane (9). The
results are
$$^{(N-1)}S^{\,\,\,\mu\nu}_{\rho}=^{(N)}S^{\,\,\,\mu\nu}_{\rho}+$$
$$\Big[n^{\mu}n_{\rho}n_{i}n^{k}e_{j}^{\nu}+n^{k}n_{\rho}n^{\nu}n_{j}e_{i}^{\mu}+$$
$$n^{\mu}n_{i}n^{\nu}n_{j}e_{\rho}^{k}-n^{k}e_{i}^{\mu}n_{\rho}e_{j}^{\nu}-n^{\mu}n_{\rho}e_{j}^{k}e_{j}^{\nu}$$
\begin{equation}
-n^{\mu}n_{\rho}n_{i}n^{k}n^{\nu}n_{j}\Big]\,^{(N)}S^{\,\,\,ij}_{k}
\end{equation}
\\

$$^{(N-1)}T^{\rho}_{\,\,\,\mu\lambda}\,^{(N-1)}S^{\,\,\,\nu\mu}_{\rho}=^{(N)}T^{\rho}_{\,\,\,\mu\lambda}\,^{(N)}S^{\,\,\,\nu\mu}_{\rho}+$$
\begin{equation}
\Big[n^{k}n_{\lambda}n^{\nu}n_{j}-e_{\lambda}^{k}n^{\nu}n_{j}-e_{\nu}^{j}n^{k}n_{\lambda}\Big]\,^{(N)}T^{m}_{\,\,\,ik}\,^{(N)}S^{\,\,\,ji}_{m}
\end{equation}
and
\begin{equation}
^{(N-1)}T=^{(N)}T
\end{equation}
The left hand side of the teleparallel field equation  which we denote by
$F_{l}^{\nu}$, can now be constructed by means of these relations
\begin{equation}
F_{l}^{\nu}=e^{-1}\partial_{\mu}(ee_{l}^{\rho}S^{\,\,\,\,\mu\nu}_{\rho})
-e_{l}^{\lambda}T_{\,\,\,\,\mu\lambda}^{\rho}S^{\,\,\,\,\nu\mu}_{\rho}-\frac{1}{4}e_{l}^{\nu}T
\end{equation}
Combining (18),(19),(20) and (21) we have
\begin{eqnarray}
\nonumber ^{(4)}F_{l}^{\nu}&=&\Bigg
[e^{-1}\partial_{\mu}(ee_{l}^{\rho}\,^{(5)}S^{\,\,\,\mu\nu}_{\rho})-e_{l}^{\lambda}
\,^{(5)}T^{\rho}_{\,\,\,\mu\lambda}\,^{(5)}S^{\,\,\,\nu\mu}_{\rho}\\
\nonumber&+&\frac{1}{4}e_{l}^{\rho}\,^{(5)}T\Bigg]+\partial_{\mu}(n^{\rho}n_{l}
\,^{(5)}S^{\,\,\,\mu\nu}_{\rho})\\
\nonumber&-&n^{\lambda}n_{l}\,^{(5)}T^{\rho}_{\,\,\,\mu\lambda}
\,^{(5)}S^{\,\,\,\nu\mu}_{\rho}\\
\nonumber&+&\frac{1}{4}n^{\nu}n_{l}\,^{(5)}T+h_{l}^{\rho}\partial_{\mu}\Big[A_{\rho
ij}^{\mu k \nu}\,^{(5)}S^{\,\,\,ij}_{k}\Big]\\
&+&A_{\rho ij}^{\mu
k\nu}\,^{(5)}S^{\,\,\,ij}_{k}\partial_{\mu}(hh_{l}^{\rho})\\
\nonumber&-&h_{l}^{\lambda}B_{\lambda
j}^{k\nu}\,^{(5)}T^{m}_{\,\,\,ik}\,^{(5)}S^{\,\,\,ji}_{m}
\end{eqnarray}
where we have defined
\begin{eqnarray}
\nonumber A_{\rho ij}^{\mu k
\nu}&=&n^{\mu}n_{\rho}n_{i}n^{k}e_{j}^{\nu}+n^{k}n_{\rho}n^{\nu}n_{j}e_{i}^{\mu}\\
\nonumber&+&n^{\mu}n_{i}n^{\nu}n_{j}e_{\rho}^{k}-n^{k}e_{i}^{\mu}n_{\rho}e_{j}^{\nu}\\
&-&n^{\mu}n_{\rho}e_{j}^{k}e_{j}^{\nu}-n^{\mu}n_{\rho}n_{i}n^{k}n^{\nu}n_{j}
\end{eqnarray}
and
\begin{equation}
B_{\lambda
j}^{k\nu}=n^{k}n_{\lambda}n^{\nu}n_{j}-e_{\lambda}^{k}n^{\nu}n_{j}-e_{\nu}^{j}n^{k}n_{\lambda}\,.
\end{equation}

\section{Junction Conditions}

In the geometrical setup of the Randall-Sundrum II model, the brane
acts as a boundary hypersurface that connects the two ``sides'' of
the bulk. In order to derive the induced field equations on the
brane, it is necessary to know how the physical quantities change
from one side of the brane to the other. This needs the so called
junction conditions in order to deal with the discontinuities across
the hypersurface. This problem has been discussed in the context of
Einstein-Cartan manifolds where both curvature and torsion are
present \cite{daSilva}. In this section we will derive the junction
conditions in our braneworld teleparallel gravity setup.\\
The derivation is conducted in the Gaussian Normal Coordinates
(GNC). In the context of general relativity, GNC is constructed as
follows: On a given hypersurface $\Sigma$, the unique geodesic is
constructed with tangent vector $n^{a}$ through each point of
$\Sigma$. Then each point in a neighborhood of $\Sigma$ is labeled
by the affine parameter $y$ along the geodesic on which it lies and
also by the arbitrary coordinates $(x_{1}; ... ; x_{n-1})$ of the
point $p\in\Sigma$ from which the geodesic originated. Then $(x_{1};
... ; x_{n-1}; y)$ defines the GNC system. This relation
$n_{a}dx^{a} = dy$ then holds and the metric takes the form
\begin{equation}
ds^{2}=g_{ab}dx^{a}dx^{b}=q_{\mu\nu}dx^{\mu}dx^{\nu}+dy^{2}
\end{equation}
The brane now can be chosen to be located at $y=0$ without any loss of
generality. A similar procedure can be applied in teleparallel
gravity. Instead of using the geodesics to define the GNC coordinate
system, we will use the teleparallel force equation which is
equivalent to the geodesic equation of general relativity. Then the
tetrad in the Gaussian normal coordinates of teleparallel gravity
takes the form
\begin{equation}
ds^{2}=\eta_{ij}h_{\mu}^{i}h_{\nu}^{j}dx^{\mu}dx^{\nu}+dy^{2}
\end{equation}
This can be justified as follows. Similar to general relativity,
with a metric in the form of (26), we can choose the unit normal
vector $n_{\mu}$ to be purely in the direction of the extra
dimension $n_{\mu}=(0,0,0,0,1)$. On the other hand, the counterpart
of $n_{\mu}$ in the tangent space is denoted by $n_{i}$. At any
point on the hypersurface (brane), the normal to the hypersurface is
also normal to the tangent space at that point. This leads us to
conclude that $n_{i}$ should also be $n_{i}=(0,0,0,0,1)$, so for the
$55$ component of the vierbein we have $e^{5}_{5}=1$ and for the
$\mu5$ components we have $e_{5}^{\mu}=0$.

In order to derive the junction conditions, we employ the language
of distributions \cite{Poisson}. Heaviside distribution $\Theta(y)$
is defined as follows: it is equal to $+1$ if $y>0$, $0$ if $y<0$
and indeterminate if $y=0$. It has the following properties
\begin{equation}
\Theta^{2}(y)=\Theta(y) \,\,\,\,\ ,
\,\,\,\,\Theta(y)\Theta(-y)=0\,\,\,\,,\,\,\,\,\frac{d}{dy}\Theta(y)=\delta(y)
\end{equation}
where $\delta(y)$ is the Dirac distribution. With this definitions,
the tetrad field can be written in the following
way
\begin{equation}
e_{\mu}^{i}=\Theta(y)e_{\mu}^{i\,\,(+)}+\Theta(-y)e_{\mu}^{i\,\,
(-)}
\end{equation}
where $e_{\mu}^{i\,\,(+)}$ and $e_{\mu}^{i\,\, (-)}$ denote the
vierbeins in the $y>0$ and $y<0$ side of the brane respectively. The
teleparallel connection, the Weitzenb\"{o}ck connection includes the
first derivative of the vierbein. Differentiating (29) and denoting
the jump of any tensorial property $f$ across the brane by
\begin{equation}
[f]\equiv f^{+}-f^{-}
\end{equation}
we have
\begin{equation}
\partial_{\rho}e_{\mu}^{i}=\Theta(y)\partial_{\rho}e_{\mu}^{i\,\,(+)}+\Theta(-y)\partial_{\rho}e_{\mu}^{i\,\,
(-)}+\delta(y)[e_{\mu}^{i}]n_{\rho}
\end{equation}
and
\begin{eqnarray}
\nonumber\Gamma_{\,\,\,\,\mu\nu}^{\rho}&=&\Theta(y)\Gamma_{\,\,\,\,\mu\nu}^{\rho\,\,(+)}+\Theta(-y)\Gamma_{\,\,\,\,\mu\nu}^{\rho\,\,(-)}+
\Theta(y)\delta(y)e_{i}^{\rho\,\,(+)}[e_{\mu}^{i}]n_{\nu}\\
&+&\Theta(-y)\delta(y)e_{i}^{\rho\,\,(-)}[e_{\mu}^{i}]n_{\nu}\,.
\end{eqnarray}
If the last two terms remain, then the Weitzenb\"{o}ck connection
will include $\Theta(y)\delta(y)$ terms which is not defined as a
distribution; so the connection itself can not be written as a
distribution. This leads to an ill-defined geometry. In order to
avoid this we impose the condition
\begin{equation}
[e_{\mu}^{i}]=0
\end{equation}
which if written in a coordinate independent way implies that
\begin{equation}
[h_{\mu}^{i}]=0
\end{equation}
The equation above is the first junction condition. This means that
the induced vierbein should be the same on both sides of the brane.

Now the connection can be written as
\begin{equation}
\Gamma_{\,\,\,\,\mu\nu}^{\rho}=\Theta(y)\Gamma_{\,\,\,\,\mu\nu}^{\rho\,\,(+)}+\Theta(-y)\Gamma_{\,\,\,\,\mu\nu}^{\rho\,\,(-)}
\end{equation}
So the torsion tensor becomes
\begin{eqnarray}
\nonumber T_{\,\,\,\,\mu\nu}^{\rho}&\equiv&\Gamma_{\,\,\,\,\mu\nu}^{\rho}-\Gamma_{\,\,\,\,\nu\mu}^{\rho}\\
&=&\nonumber
\Theta(y)\Gamma_{\,\,\,\,\mu\nu}^{\rho\,\,(+)}+\Theta(-y)\Gamma_{\,\,\,\,\mu\nu}^{\rho\,\,(-)}\\
\nonumber&-&\Theta(y)\Gamma_{\,\,\,\,\nu\mu}^{\rho\,\,(+)}+\Theta(-y)\Gamma_{\,\,\,\,\nu\mu}^{\rho\,\,(-)}\\
&=&\Theta(y)T_{\,\,\,\,\mu\nu}^{\rho\,\,(+)}+\Theta(-y)T_{\,\,\,\,\mu\nu}^{\rho\,\,(-)}
\end{eqnarray}
This means that the torsion tensor can be written as a distribution
and it has no $\delta$-term. This is in contrast to the general relativity
where the Riemann tensor includes a $\delta$-term. Consequently the
superpotential and the torsion scalar also can be written as
distributions
\begin{equation}
S^{\,\,\,\,\mu\nu}_{\rho}=\Theta(y)S^{\,\,\,\,\mu\nu\,\,(+)}_{\rho}+\Theta(-y)S^{\,\,\,\,\mu\nu\,\,(-)}_{\rho}
\end{equation}
and
\begin{equation}
T=S^{\,\,\,\,\mu\nu}_{\rho}T_{\,\,\,\,\mu\nu}^{\rho}=\Theta(y)T^{(+)}+\Theta(-y)T^{(-)}
\end{equation}

Now we turn our attention to the 5-dimensional teleparallel field
equations
\begin{equation}
e^{-1}\partial_{\mu}(ee_{l}^{\rho}S^{\,\,\,\,\mu\nu}_{\rho})-e_{l}^{\lambda}T_{\,\,\,\,\mu\lambda}
^{\rho}S^{\,\,\,\,\nu\mu}_{\rho}-\frac{1}{4}e_{l}^{\nu}T
=4\pi G e_{l}^{\rho} \Xi_{\rho}^{\nu}
\end{equation}
where $\Xi_{\rho}^{\nu}$ is the $5$-dimensional energy-momentum
tensor. Any discontinuity ($\delta$-term) in the left hand side of
the above equation should be related to the matter energy-momentum
tensor on the brane via the second junction condition which we seek
to derive here. From equations (36) , (37) and (38) and the form of
the field equations, it is obvious that the only term which is
capable of producing any $\delta$-term and discontinuity in (39) is
the first term while the other two terms have no $\delta$-term.
Using (37) we'll have
\begin{eqnarray}
\nonumber\partial_{\mu}(S^{\,\,\,\,\mu\nu}_{\rho})&=&\Theta(y)\partial_{\mu}(S^{\,\,\,\,\mu\nu\,\,(+)}_{\rho})
+\Theta(-y)\partial_{\mu}(S^{\,\,\,\,\mu\nu\,\,(-)}_{\rho})\\
&+&\delta(y)[S^{\,\,\,\,\mu\nu}_{\rho}]n_{\mu}
\end{eqnarray}
Assuming the bulk to be empty except for a cosmological constant
$\Lambda_{5}$, the 5-dimensional energy-momentum tensor can be
decomposed as
\begin{equation}
\Xi_{i}^{\nu}=-\Lambda_{5}\, e_{i}^{\nu}+\delta(y)\,\Omega_{i}^{\nu}
\end{equation}
where $\Omega_{i}^{\nu}$ is the matter energy-momentum tensor on the
brane. Equalling the $\delta$-terms on the two sides of the
teleparallel field equation (39), we get
\begin{equation}
e_{i}^{\rho}\,[S^{\,\,\,\,\mu\nu}_{\rho}]\,n_{\mu}=4\pi
G\,\,\Omega_{i}^{\nu}
\end{equation}
This is the second junction condition. It implies that the jump in
the superpotential tensor across the brane is related to the matter
energy-momentum tensor confined to the brane.

\section{Induced Teleparallel Field Equations on the Brane}

All the elements required to derive the induced teleparallel field
equations on the brane are now prepared. The quantities can be
evaluated on either side of the brane by imposing the
$Z_{2}$-symmetry. Using the junction condition (42) to relate
$S^{\,\,\,\,\mu\nu}_{\rho}\,n_{\mu}$ terms in (23) to the matter
energy-momentum on the brane , we have
\begin{equation}
^{(4)}F_{l}^{\nu}=-\Lambda_{5}h_{l}^{\nu}+(4\pi
G_{5})^{2}\Pi_{l}^{\nu}+E_{l}^{\nu}
\end{equation}
where we have defined
\begin{eqnarray}
\nonumber\Pi_{l}^{\nu}=&-&\frac{3}{4}h_{\rho}^{i}\Omega_{i}^{\nu}\Omega^{\rho}_{l}+\frac{3}{8}h_{l}^{\rho}\Omega\Omega_{\rho}^{\nu}\\
\nonumber&+&\frac{1}{32}h_{l}^{\nu}\Omega_{i}^{\rho}\Omega^{i}_{\rho}+\frac{1}{32}h_{l}^{\nu}\Omega^{2}\\
&+&\frac{1}{4}\delta_{\rho}^{\nu}\Omega\Omega_{l}^{\rho}
+\frac{1}{4}\delta_{l}^{\nu}\Omega^{2}
\end{eqnarray}
and
\begin{eqnarray}
E_{l}^{\nu}&=&n^{\rho}n_{l}\partial_{\mu}(S^{\,\,\,\mu\nu}_{\rho})+S^{\,\,\,\mu\nu}_{\rho}(n^{\rho}\partial_{\mu}n_{l})\\
\nonumber&+&S^{\,\,\,\mu\nu}_{\rho}(n^{l}\partial_{\mu}n_{\rho})+h_{l}^{\rho}S^{\,\,\,\mu\nu}_{\rho}(n^{i}\partial_{\mu}n_{i})\\
\nonumber&+&\Bigg[n^{\mu}n_{\rho}n_{i}n^{k}e_{j}^{k}+n^{\nu}n_{\rho}n_{j}n^{k}e_{i}^{\mu}+n^{\mu}n_{i}n_{j}n^{\nu}e_{\rho}^{k}\\
\nonumber&-&n^{k}n_{\rho}e_{i}^{\mu}e_{j}^{\nu}-n^{\mu}n_{i}e_{\rho}^{k}e_{j}^{\nu}-n^{\mu}n_{\rho}n_{i}n^{k}n^{\nu}n_{j}\Bigg]
S^{\,\,\,ij}_{k}\partial_{\mu}(hh_{l}^{\rho})
\end{eqnarray}
By further decomposing the brane stress-energy tensor into the brane
tension and energy-momentum tensor of the matter
\begin{equation}
\Omega_{l}^{\nu}=-\lambda h_{l}^{\nu}+\tau_{l}^{\nu}
\end{equation}
where $\lambda$ is the tension of the brane in 5 dimensions and
$\tau_{l}^{\nu}$ is the energy-momentum tensor, the equation (43)
can be written as
\begin{equation}
^{(4)}F_{l}^{\nu}=-\Lambda_{4}h_{l}^{\nu}+4\pi
G_{4}\tau_{l}^{\nu}+(4\pi G_{5})^{2}\pi_{l}^{\nu}+E_{l}^{\nu}
\end{equation}
where
\begin{equation}
\Lambda_{4}=\Lambda_{5}+(4\pi G_{5})^{2}\lambda^{2}
\end{equation}
and
\begin{equation}
G_{4}=\frac{(4\pi G_{5})^{2}\lambda }{3\pi}
\end{equation}
and $\lambda$ is the brane tension and $h_{l}^{\nu}$ is the induced
vierbein.

We also have
\begin{equation}
\pi_{l}^{\nu}=-\frac{3}{4}\tau_{\rho}^{\nu}\tau_{l}^{\rho}+\frac{3}{8}\tau\tau_{l}^{\nu}+\frac{3}{8}h_{l}^{\nu}\tau_{i}^{\rho}\tau_{\rho}^{i}
+\frac{3}{16}h_{l}^{\nu}\tau^{2}+\frac{1}{4}\delta_{l}^{\nu}\tau_{i}^{\rho}\tau_{\rho}^{i}+\frac{1}{4}\delta_{l}^{\nu}\tau^{2}
\end{equation}
Equation (47) is the induced teleparallel field equation on the
brane. The last  two terms in (50) are not present in GR version of
field equations. They are extra terms arising from extra degrees of
freedom in teleparallel gravity. By conducting the calculation in
the Gaussian Normal coordinates(GNC), the $E_{l}^{\nu}$ term that
carries the effects of the bulk geometry on the brane, will be
greatly simplified. In this coordinate system the only non-zero
terms of $E_{l}^{\nu}$ are:
\begin{equation}
E_{l}^{\nu}=n_{l}n^{\rho}\partial_{\mu}(S^{\,\,\,\mu\nu}_{\rho})+\frac{5}{2}T_{\,\,\,\alpha}^{\alpha\nu}n_{l}
\end{equation}
where $T^{\,\,\,\mu\nu}_{\rho}$ is the bulk torsion tensor and the
normal vector is purely in the direction of the extra dimension
(only its 5th component is nonzero). The torsion tensor can be
decomposed into three components \cite{CLS}, namely a vector part:
\begin{equation}
V_{\mu}=T_{\,\,\,\nu\mu}^{\nu}
\end{equation}
an axial part:
\begin{equation}
A^{\mu}=\frac{1}{6}\epsilon^{\mu\nu\rho\sigma}T_{\nu\rho\sigma}
\end{equation}
and a purely tensorial part:
\begin{equation}
P_{\lambda\mu\nu}=\frac{1}{2}(T_{\lambda\mu\nu}+T_{\mu\lambda\nu})+\frac{1}{6}(g_{\nu\lambda}V_{\mu}
+g_{\nu\mu}V_{\lambda})-\frac{1}{3}g_{\lambda\mu}V_{\nu}
\end{equation}
it is obvious from equation $(51)$ that the  $E_{l}^{\nu}$ consists
only the vector part of the bulk torsion tensor. It may be useful to
remember that in GR the corresponding term is given by the
projection of the bulk Weyl tensor into the brane. In the
cosmological context, when the Friedmann equation come to account,
the corresponding term in GR results in the so called dark radiation
term. This term plays a crucial role in studying the dynamics of
cosmological perturbations. The role of this term in teleparallel
gravity is currently under investigation by the present authors.

\section{Conclusion and Discussion}
It is a well known fact that teleparallel theory and its extensions
are not locally Lorentz invariant. As a result in four dimensions,
there are 10 independent components of the metric but 16 independent
component of the tetrad. Also there are infinite number of tetrads
that can be chosen to construct any given metric. Each of these
tetrads are related through a Lorentz transformation (or rotation)
in the tangent space. In this paper using a projection vierbein in
the form of (12), we have derived the equivalent of the
Gauss-Codacci equations in brane teleparallel gravity. Similar to
general relativity, any discontinuity in the geometrical part of the
field equation should be related to the brane stress-energy content
through junction conditions. In general relativity these junction
conditions are given in terms of the jump in the extrinsic curvature
across the brane. However in teleparallel gravity the jump in the
normal part of the superpotential tensor is related to the matter on
the brane (equation (42)). Using these relations, we derived the
induced teleparallel field equation on the brane (equation (47)).
Compared to general relativity there are two extra terms in the
teleparallel version of the field equation corresponding to extra
degrees of freedom in the theory. In general relativity the term
carrying the effects of the bulk on the brane is given by the
projection of the bulk Weyl tensor on the brane. However in our
brane teleparallel gravity model this term is given by equation
(51). This means that the corresponding term in teleparallel gravity
consists of the projection of the vector part of the bulk torsion
tensor on the brane.

\appendix*
\section{}
In this appendix we present the detailed proof of the Gauss equation
in teleparallel gravity, equation (18). From the definition of the
covariant derivative on the brane, equation (14), we have for any
arbitrary vector $X_{\rho}$
\begin{eqnarray}
\nonumber
D_{\mu}D_{\nu}X_{\rho}&=&h_{\nu}^{l}h_{\rho}^{m}h_{\mu}^{k}h_{l}^{\alpha}h_{m}^{\beta}
\nabla_{k}\Big[h_{\alpha}^{i}h_{\beta}^{j}\nabla_{i}(e_{j}^{\gamma}X_{j})\Big]\\
\nonumber&=&h_{\nu}^{l}h_{\rho}^{m}h_{\mu}^{k}h_{l}^{\alpha}h_{m}^{\beta}h_{\alpha}^{i}h_{\beta}^{j}\nabla_{k}\nabla_{i}(e_{j}^{\gamma}X_{j})\\
&+&h_{\nu}^{l}h_{\rho}^{m}h_{\mu}^{k}h_{l}^{\alpha}h_{m}^{\beta}\nabla_{k}(h_{\alpha}^{i}h_{\beta}^{j})\nabla_{i}(e_{j}^{\gamma}X_{j})
\end{eqnarray}
Swapping $\mu$ and $\nu$ and subtracting, we have from equation (17)
\begin{eqnarray}
\nonumber
^{(N-1)}T^{\delta}_{\,\,\,\mu\nu}D_{\delta}X_{\rho}&=&h_{\nu}^{l}h_{\rho}^{m}h_{\mu}^{k}h_{l}^{\alpha}h_{m}^{\beta}h_{\alpha}^{i}h_{\beta}^{j}
\nabla_{k}\nabla_{i}(e_{j}^{\gamma}X_{j})\\
\nonumber&+&h_{\nu}^{l}h_{\rho}^{m}h_{\mu}^{k}h_{l}^{\alpha}h_{m}^{\beta}\nabla_{k}(h_{\alpha}^{i}h_{\beta}^{j})\nabla_{i}(e_{j}^{\gamma}X_{j})\\
\nonumber&-&h_{\mu}^{l}h_{\rho}^{m}h_{\nu}^{k}h_{l}^{\alpha}h_{m}^{\beta}h_{\alpha}^{i}h_{\beta}^{j}\nabla_{i}\nabla_{k}(e_{j}^{\gamma}X_{j})\\
\nonumber&-&h_{\mu}^{l}h_{\rho}^{m}h_{\nu}^{k}h_{l}^{\alpha}h_{m}^{\beta}\nabla_{i}(h_{\alpha}^{i}h_{\beta}^{j})\nabla_{k}(e_{j}^{\gamma}X_{j})\\
\end{eqnarray}
We now use the definition of the projection vierbein (12) and the
fact that the Weitzenb\"{o}ck covariant derivative of the tetrad
field is zero. The result after some manipulation is
\begin{eqnarray}
\nonumber^{(N-1)}T^{\delta}_{\,\,\,\mu\nu}D_{\delta}X_{\rho}=h_{\nu}^{i}h_{\rho}^{j}h_{k}^{\mu}e_{j}^{\gamma}\,
^{(N)}T^{\alpha}_{\,\,\,ki}\nabla_{\alpha}X_{\gamma}\\
\nonumber+e_{j}^{\gamma}h_{\nu}^{l}h_{\rho}^{m}h_{\mu}^{k}h_{l}^{\alpha}h_{m}^{\beta}
\nabla_{k}\Big[(e_{\alpha}^{i}-n^{i}n_{\alpha})(e_{\beta}^{j}-n^{j}n_{\beta})\Big]\nabla_{i}X_{j}\\
\nonumber-e_{j}^{\gamma}h_{\mu}^{l}h_{\rho}^{m}h_{\nu}^{k}h_{l}^{\alpha}h_{m}^{\beta}\nabla_{i}(h_{\alpha}^{i}h_{\beta}^{j})\nabla_{k}X_{j}\\
\end{eqnarray}
We now develop the R.H.S of the above equation
\begin{equation}
^{(N-1)}T^{\delta}_{\,\,\,\mu\nu}D_{\delta}X_{\rho}=^{(N-1)}T^{\delta}_{\,\,\,\mu\nu}
e_{i}^{\gamma}h_{\delta}^{j}h_{\rho}^{i}e_{j}^{\alpha}\nabla_{\alpha}X_{\gamma}
\end{equation}
Combining the results and noting that
$h_{\mu}^{i}n^{\mu}=h_{\nu}^{j}n_{j}=0$, we'll have
\begin{equation}
^{(N-1)}T^{\delta}_{\,\,\,\mu\nu}
e_{i}^{\gamma}h_{\delta}^{j}h_{\rho}^{i}e_{j}^{\alpha}\nabla_{\alpha}X_{\gamma}=h_{\nu}^{i}h_{\rho}^{j}h_{k}^{\mu}e_{j}^{\gamma}\,
^{(N)}T^{\alpha}_{\,\,\,ki}\nabla_{\alpha}X_{\gamma}
\end{equation}
Removing the arbitrary vector and rearranging, we get the desired
result.
\begin{equation}
^{(N-1)}T^{\rho}_{\,\,\,\mu\nu}=h_{k}^{\rho}h_{\nu}^{j}h_{\mu}^{i}\,
^{(N)}T^{k}_{\,\,\,ij}
\end{equation}

\end{document}